\documentclass[showpacs,a4paper,twocolumn,10pt]{revtex4-1}
\usepackage[utf8x]{inputenc}
\usepackage{amsmath}
\usepackage{amsfonts}
\usepackage{amssymb}
\usepackage{graphicx}

\usepackage{amsmath}
\usepackage{amsfonts}
\usepackage{amssymb}
\usepackage{amsthm}
\usepackage{mathrsfs}

	\usepackage{catchfile}
	\newcommand{\getenv}[2][]{%
		\CatchFileEdef{\temp}{"|kpsewhich --var-value #2"}{}%
		\if\relax\detokenize{#1}\relax\temp\else\let#1\temp\fi%
	}
	\getenv[\SHR]{SHR}
	\getenv[\PTEN]{P10}

	\DeclareMathAlphabet{\mathpzc}{OT1}{pzc}{m}{it}

	\newcommand{\mr}[1]{\mathrm{#1}}			

	\newcommand{\br}[1]{\left( #1 \right)}
	\newcommand{\brr}[1]{\left[ #1 \right]}
	\newcommand{\brrr}[1]{\left\{ #1 \right\}}
	\newcommand{\of}[1]{\hspace{-2.5pt}\br{#1}}

		\newcommand{\Sum}[2]{\sum\limits_{#1}^{#2}}

		\newcommand{\Int}[3]{\int\limits_{#1}^{#2}\mr{d}#3\,}
		\newcommand{\sInt}[3]{\int_{#1}^{#2}\mr{d}#3\,}

	\renewcommand{\d}{\mathrm{d}}

	\newcommand{\Landau}[1]{\mathpzc{O}\left( #1 \right)}
	\newcommand{\landau}[1]{\mathpzc{o}\left( #1 \right)}
 
		\newcommand{\Min}[2]{\min\of{#1,#2}}
		
		\newcommand{\Abs}[1]{\left\vert #1 \right\vert}

		\newcommand{\Gma}[1]{\Gamma\of{#1}}

		\newcommand{\TA}[1]{\overline{#1}}
		\newcommand{\xpc}[1]{\left\langle #1 \right\rangle}
		\newcommand{\sxpc}[1]{\langle #1 \rangle}

\begin{document}

	\title{Weak ergodicity breaking in an anomalous diffusion process of mixed origins}

	\author{Felix Thiel}
	\author{Igor M. Sokolov}\affiliation{Institute of Physics, Humboldt
	  University Berlin, Newtonstr. 15, 12489 Berlin, Germany}
	\date{\today}

	\begin{abstract}
		The ergodicity breaking parameter is a measure for the heterogeneity among different trajectories of one ensemble.
		In this report this parameter is calculated for fractional Brownian motion with a random change of time scale, often called ``subordination''.
		We proceed to show that this quantity is the same as the known CTRW case.
	\end{abstract}
	\maketitle

		``Weak ergodicity breaking'' is a term which occurred in the spotlight a few years ago, see for instance \cite{Brokmann2003,Jeon2010,He2008,Deng2009,Rebenshtok2007,Cherstvy2013} with respect to observables showing aging behavior.
		The term ``aging'' applies to the specific behavior pertinent to the values of an observable measured at two distinct instants of time, like the mean squared displacement (MSD) in the time interval between $t_1$ and $t_2$, $\xpc{ \Delta X^2\of{t_1,t_2} }$ as depending on the age $t_1$ of the system (assumed to be prepared at $t=0$) at the beginning of observation.
		Aging here essentially refers to a non-stationarity of the observed process, in which $\xpc{\Delta X^2\of{t_1,t_2} }$  cannot be expressed via the time lag $\tau = t_2 - t_1$ only and keeps a considerable dependence on $t_1$ as a stand-alone variable.

		Let us concentrate now on the ergodicity properties the squared displacement during the time interval $\tau$ in a measurement starting at $t$.
		In the case when the data acquisition in each single run of the process takes time from $t=0$ to $t=T$, the time average of $\Delta X^2\of{t,t+\tau}$ can be performed:
		\begin{equation}
				\TA{\Delta X^2\of{\tau}} 
			= 
				\lim_{T \to \infty} \frac{1}{T - \tau} \sInt{0}{T-\tau}{t} \Delta X^2\of{t,t+\tau}
			.
			\label{eqERG}
		\end{equation}
		The process is considered ergodic provided $\TA{\Delta X^2\of{\tau}} = \xpc{ \Delta X^2\of{t,t+\tau}}$.
		The discussion of the ergodicity implies that the corresponding limit does exist in some probabilistic sense (say, in probability) and is equal to the ensemble mean.
		Note that the time-average over the data acquisition interval removes the explicit $t$-dependence.
		For non-stationary processes the ensemble mean at any $t$ depends on $t$, while the time-integration over the data acquisition interval removes this dependence.
		Therefore, even provided the integral above converges in a whatever sense, it cannot converge to all $\xpc{\Delta X^2\of{t,t+\tau}}$ simultaneously, and our process is trivially non-ergodic.

		Whether the random process is stationary or not, one can ask how diverse are its different realizations, i.e. how different are two trajectories of the process with respect to some time-averaged observable $O\of{\tau;t}$ (in our case the MSD, $O\of{\tau;t} = \Delta X^2\of{t,t+\tau}$).
		To this purpose one considers a fixed-$T$ approximation to eq.\eqref{eqERG}
		\begin{equation*}
				\TA{O\of{\tau}}_T
			=
				\frac{1}{T-\tau} \sInt{0}{T-\tau}{t} O\of{\tau;t}
			.
		\end{equation*}
		At any finite $T$ the value of $\TA{O\of{\tau}}_T$ is a random variable.
		For both, stationary and non-stationary processes $O(\tau,t)$, one can consider a parameter describing the strength of fluctuations of $\TA{O\of{\tau}}_T$.
		As a measure of homogeneity or heterogeneity of different trajectories one can take the relative dispersion of the $\TA{O\of{\tau}}_T$ in different realizations of the process:
		\begin{equation}
				J\of{\tau,T} 
			= 
				\frac{
					\sxpc{\TA{O\of{\tau}}^2_T} - \sxpc{\TA{O\of{\tau}}_T}^2
				}{
					\sxpc{\TA{O\of{\tau}}_T}^2
				}
			.
			\label{eqJPar}
		\end{equation}
		The parameter $J$ (often called ``ergodicity breaking parameter'') shows how different are different trajectories of the process with respect to the observable $O$.
		For stationary processes, vanishing of $J$ in the long time limit indeed implies ergodicity \cite{Rytov1987}, and its non-vanishing witnesses against ergodicity of the process.
		If for any fixed $T$ this relative dispersion stays finite or diverges, this may be considered as a nonexistence of the limit in eq.\eqref{eqERG}).

		The existence of the finite limit $J = \lim_{T \to \infty} J\of{\tau,T}$ hints onto universal fluctuation behavior, as it is the case in continuous time random walks (CTRW) \cite{He2008}.
		Similar behavior (often called ``weak ergodicity breaking'') was observed in some other theoretical models \cite{He2008,Deng2009,Jeon2010,Rebenshtok2007,Lomholt2007,Heinsalu2006,Margolin2005}, as well as experimentally in blinking quantum dots \cite{Brokmann2003} and in the diffusion of ion channels on the membrane of living cells \cite{Weigel2011}.
		The name ``weak ergodicity breaking'' is slightly misleading, since this behavior may be missing in some trivially non-ergodic processes \cite{Thiel2013a} still showing $J=0$.
		We therefore prefer to call $J$ the ``heterogeneity parameter'', or ``ensemble heterogeneity''.

		In what follows we consider ensemble heterogeneity in a model for anomalous diffusion of mixed origin.
		Anomalous diffusion is a random process whose MSD shows a power-law time-dependence $\xpc{\Delta X^2\of{0,\tau}} = 2 K_{\alpha} \tau^{\alpha}$, with some $\alpha \neq 1$ (usually between zero and two).
		In the following text, we choose our units such, that $2 K_{\alpha} = 1$.
		Many different mathematical models (like CTRW, fractional Brownian motion, diffusion on fractal substrates, etc.) with different physical mechanisms have been used successfully to explain the algebraic growth of the MSD, see \cite{Sokolov2012}.
		However, in the last years it became clear that the observed anomaly may be of mixed origins \cite{Meroz2010,Sokolov2012} (see \cite{Weigel2011,Jeon2011} for related experiments), i.e. it arises from the combination of different mechanisms.
		This entanglement of different mechanism poses a new challenge to the theorist and raises the question, how the process may be decomposed and understood \cite{Thiel2013}.

		In this report we will consider a simple process of mixed origin, a subordinated fractional Brownian motion (sfBm), and calculate its heterogeneity parameter.
		Special cases of this problem have already been solved by Barkai, Metzler and coworkers \cite{He2008,Deng2009}, but not in its full complexity.
		We show, that the heterogeneity parameter assumes its CTRW value (i.e. is governed by the leading process alone) in all valid situations.
		Therefore it is identified as a property inherited from the non-stationary part of the process.

		The rest of the paper is organized as follows: first we will refine the definition of $J$, then define the process in question.
		After the calculations are carried out, we will discuss the result.

	\textbf{The heterogeneity parameter:}
		The observable in question is the time averaged MSD (TA MSD).
		Let us consider therefore a one dimensional random process $X\of{t}$ measured on the time interval $\brr{0,T}$.

		Denote the time lag  $0<\tau< T/2$ and concentrate on the increments $\Delta X_{t} = X\of{t+\tau} - X\of{t}$.
		The TA MSD reads
		\begin{equation}
				\TA{\Delta X^2\of{\tau}}_T
			= 
				\frac{1}{T-\tau} \sInt{0}{T-\tau}{t} \Delta X^{2}_{t}
			.
			\label{TAMSDDef}
		\end{equation}

		Plugging this into eq.\eqref{eqJPar}, we obtain the ensemble heterogeneity:
		\begin{equation}
				J\of{\tau,T} 
			= 
				\frac{
					\Int{0}{T-\tau}{t_1} \Int{0}{T-\tau}{t_2}  
					\xpc{ \Delta X^{2}_{t_1} \Delta X^{2}_{t_2} } 
				}{ 
					\xpc{ \sInt{0}{T-\tau}{t}  \Delta X^2_{t}  }^2
				} 
				- 1
			.
			\label{eqEDDef}
		\end{equation}
		By dimensional arguments, $J$ will only depend on $\tau / T$.

	\textbf{Subordinated fractional Brownian motion:}
		As an example for diffusion of mixed origins we will discuss a process arising from the subordination of fractional Brownian motion (fBm) to a renewal process with power law waiting time density: the subordinated fractional Brownian motion (sfBm).
		
		It can be described as follows.
		Let $\tilde{X}\of{u}$ be a fBm defined for $u \in [0,\infty)$, that is a Gaussian process with covariance matrix
		\begin{equation}
				\xpc{ \tilde{X}\of{u_1} \tilde{X}\of{u_2} }
			=
				\frac{1}{2} \brr{
					u_{1}^{2H} + u_{2}^{2H} - \Abs{u_1 - u_2}^{2H}
				} 
			\label{Cov}
		\end{equation}
		and zero mean.
		$H \in \br{0,1}$ is the so called Hurst or self-similarity index of the process.
		The index determines the growth of the MSD via $\langle \tilde{X}^2\of{u} \rangle = u^{2H}$.
		If $H > 1/2$, the process is superdiffusive.
		It is subdiffusive for $H < 1/2$.
		For $H = 1/2$, the process reduces to a Brownian motion.
		FBm is a model often used to describe processes in polymer dynamics \cite{Walter2012,Zoia2009} and diffusion in crowded media like biological cells \cite{Weber2010,Szymanski2009,Burnecki2012}.
		The physical mechanism underlying fBm is a memory feedback between the diffusing particle and its environment, which introduces correlations that either enhance or impair diffusion, as it is the case for a particle interacting with its viscoelastic medium, or for a monomer being coupled to the rest of a polymer chain.

		Let $U\of{t}$ be a renewal process with non-negative increments and $t \in \brr{0,T}$.
		Here we consider a counting process with unit increments, following after waiting times distributed according to a one-sided $\alpha$-stable distribution, with $\alpha \in \br{0,1}$.

		SfBm is the composition of both processes, i.e.
		\begin{equation*}
			X\of{t} := \tilde{X}\of{U\of{t}}
			\text{.}
		\end{equation*}
		This method of composition is referred to as ``subordination'' or ``time change'' in literature.
		In this language, the process $X\of{t}$ is called \textit{subordinated} to the \textit{parent process} $\tilde{X}\of{u}$ via the \textit{leading process} $U\of{t}$ \cite{Feller1968}.
		Subordinated processes find applications in finance \cite{Gu2012} and are also studied in mathematics \cite{Linde2004,Zhang2013,KumarA2011,Meerschaert2009,Hahn2011}.
		We need to stress here, that $U\of{t}$ is not the inverse of a stable motion.
		In contrast to such a process, $U\of{t}$ does neither possess independent nor stationary increments.

		SfBm with $H=1/2$ is related to continuous-time random walks (CTRW).
		A CTRW describes a particle diffusing in an energy trap landscape.
		The particle is trapped at a certain position for a long time, because it is not able to escape from the potential well.
		In that way, diffusive transport is hindered.
		Note, that $X\of{t}$ corresponds to an off-lattice CTRW with Gaussian step length distribution, when $U\of{t}$ is a renewal process with power law waiting time distribution (as in our case).
 
		The heterogeneity parameter for fBm and simple CTRW has already been calculated in \cite{Deng2009} and in \cite{He2008}.
		The result for the general composite process $X\of{t}$ is -- up to our knowledge -- new.

	\textbf{The calculation:}
		To obtain $J$, we need to find expressions for $\langle \Delta X^2_{t} \rangle$ and for $\langle \Delta X^2_{t_1} \Delta X^2_{t_2} \rangle$.
		This is easily done by calculating the conditional expectations, by first averaging over realizations of $\tilde{X}\of{u}$.
		According to eq.\eqref{Cov}
		\begin{align*}
				\xpc{ \Delta X_{t_1} \Delta X_{t_2} } 
			= &
				\xpc{ \xpc{ \Delta \tilde{X} \of{U\of{t_1}} \Delta \tilde{X}\of{U\of{t_2}} }_{\tilde{X}} }_U 		\\
			= &
				\frac{1}{2} \left\langle  
					\Abs{ \Delta U_{t_1,t_2+\tau} }^{2H} 
					+ \Abs{ \Delta U_{t_1+\tau,t_2} }^{2H}
				\right.																								\\
			& 
				\quad \left.
					- \Abs{ \Delta U_{t_1+\tau,t_2+\tau} }^{2H}
					- \Abs{ \Delta U_{t_1,t_2} }^{2H}
				\right\rangle_U
			\text{.}
		\end{align*}
		In order to calculate higher order correlators, we exploit the Gaussianity of $\tilde{X}$
		\begin{equation*}
				\xpc{ \Delta \tilde{X}^2_{u_1} \Delta \tilde{X}^2 _{u_2} }_{\tilde{X}}
			= 
				\xpc{ \Delta \tilde{X}^2_{u_1} }_{\tilde{X}} \xpc{ \Delta \tilde{X}^2_{u_2} }_{\tilde{X}}
				- 2 \xpc{ \Delta \tilde{X}_{u_1} \Delta \tilde{X}_{u_2} }_{\tilde{X}}^2
			\text{.}
		\end{equation*}
		Using these two expressions, we obtain $J$ in terms of $U$:
		\begin{align*}
				J\of{\tau,T}
			= &
				\frac{ 
					\Int{0}{T-\tau}{t_1} \Int{0}{T-\tau}{t_2}
					\xpc{ \Delta U_{t_1}^{2H} \Delta U_{t_2}^{2H} } 
					-
					\xpc{ \Delta U_{t_1}^{2H}} \xpc{ \Delta U_{t_2}^{2H} } 
				}{ 
					\xpc{ \sInt{0}{T-\tau}{t} \Delta U_{t}^{2H} }^2 
				}																										\\
			& \quad +
				\frac{1}{2} \frac{ 
					\Int{0}{T-\tau}{t_1} \Int{0}{T-\tau}{t_2}
					\xpc{ K\of{\Min{t_1}{t_2},\Abs{t_1-t_2}} }
				}{ 
					\xpc{ \sInt{0}{T-\tau}{t} \Delta U_{t}^{2H} }^2 
				}
			.
		\end{align*}
		$K\of{t_m,\tau_d}$ is an abbreviation for
		\begin{align}
			\nonumber &
				K\of{t_m,\tau_d}	 																		
			\nonumber = 
				\left[ \br{ 
					\Delta U_{t_m,t_m+\tau_d+\tau}^{2H}
					- \Delta U_{t_m,t_m+\tau_d}^{2H}
				} \right.																								\\
			& \quad -
				\left. \big(
					\Delta U_{t_m+\tau,t_m+\tau_d+\tau}^{2H}
					- \Abs{\Delta U_{t_m+\tau,t_m+\tau_d}}^{2H}
				\big) \right]^2 
				\label{CorrDef}
		\end{align}
		and owes its form to the correlation function of $\tilde{X}$, eq.\eqref{Cov}.
		We will refer to it as ``correlation part'' in the rest of this paper.
		The usual covariance in the enumerator of the first summand represents a contribution to $J$ which only depends on properties of the leading process $U$, and will be called ``covariance part'' in the following.
		Applying the variable transformation $t_m = \Min{t_1}{t_2}$ and $\tau_d = \Abs{t_1-t_2}$ and using the symmetry of the integral with respect to interchange of $t_1$ and $t_2$, we arrive at:
		\begin{align}
			\nonumber
				J\of{\tau,T}
			= &
				2 \frac{ 
					\Int{0}{T-\tau}{\tau_d} \Int{0}{T-\tau-\tau_d}{t_m}
					\xpc{\Delta U_{t_m}^{2H} \Delta U_{t_m+\tau_d}^{2H}} 
				}{ 
					\xpc{ \sInt{0}{T-\tau}{t} \Delta U_{t}^{2H} }^2 
				}																										\\
			& \quad +
				\frac{ 
					\Int{0}{T-\tau}{\tau_d} \Int{0}{T-\tau-\tau_d}{t_m}
					\xpc{ K\of{t_m,\tau_d} }
				}{ 
					\xpc{ \sInt{0}{T-\tau}{t} \Delta U_{t}^{2H} }^2 
				}																										
				- 1
			\label{eqNasty} 
			.
		\end{align}

		The simple limiting cases correspond to $H = 1/2$ (CTRW), $\alpha \rightarrow 1$ (pure fBm), both already discussed in the literature, and $H\rightarrow 1$.
		In the last case, $J$ can be calculated from eq.\eqref{eqNasty}, although the underlying fBm is only defined for $H<1$.

		Let us first consider the limit $\alpha \rightarrow 1$.
		In this regime $U\of{t}$ reduces to a Poisson process with some intensity $\lambda$.
		In the limit of long times the composition $\tilde{X}\of{U\of{t}}$ evaluates to a simple rescaling of time $X\of{t} = \tilde{X}\of{\lambda t}$.
		We can therefore continue by putting $U\of{t} = \lambda t$ and referring to the result of \cite{Deng2009}, where $J$ was calculated to be zero in the limit of diverging $T$.

		If, on the other hand, $H$ goes to unity, the correlation part, eq.\eqref{CorrDef}, equals to $4 \xpc{ \Delta U^{2}_{t_m} \Delta U^{2}_{t_m+\tau_d} }$.
		We will soon calculate the integral over such expectations.

		Last but not least of the special cases, let $H$ be one half.
		The fBm reduces to a simple Brownian motion and the correlation part vanishes identically, for all $\tau_d > \tau$.
		As mentioned above, $X\of{t}$ is an off-lattice CTRW.
		Simple CTRW was considered in \cite{He2008}, where the result
		\begin{equation}
				J_{\text{CTRW}}\of{T\rightarrow\infty}
			=
				2 \frac{ \Gamma^2\of{1+\alpha} }{ \Gma{1+2\alpha} } - 1 
		\end{equation} 
		was obtained.

		The general case leads to lengthy calculations of correlation functions of $U^{2H}$ which are given in the appendix.
		Here the results are given in the leading order of $T$:
		The denominator reads
		\begin{equation}
				\xpc{ \Int{0}{T-\tau}{t} \Delta U_{t}^{2H} }^2
			= 
				\frac{ 
					\Gamma^2\of{1+2H} \tau^{2+2\alpha\bar{H}} T^{2\alpha} 
				}{ 
					\Gamma^2\of{2+\alpha \bar{H}} \Gamma^2\of{1+\alpha} 
				} + \Landau{T^{2\alpha-1}}
			\text{,}
			\label{DenomResult}
		\end{equation}
		where $\bar{H}$ is an abbreviation for $\bar{H}=2H-1$.
		The covariance part is
		\begin{align}
			\nonumber 																		
				\Int{0}{T-\tau}{\tau_d} \Int{0}{T-\tau-\tau_d}{t_m} 
				\xpc{ \Delta U_{t_m}^{2H} \Delta U_{t_m+\tau_d}^{2H} }				\\
			= 
				\frac{
					\Gamma^2\of{1+2H} \tau^{2+2\alpha\bar{H}} T^{2\alpha}
				}{ 
					\Gma{1+2\alpha} \Gamma^2\of{2+\alpha \bar{H}} 
				} + \landau{T^{2\alpha}}
			\text{.}
			\label{CovResult}
		\end{align}
		The correlation part can be estimated as
		\begin{equation}
				\sInt{0}{T-\tau}{\tau_d} \sInt{0}{T-\tau-\tau_d}{t_m} 
				\xpc{ K\of{t_m,\tau_d} }
			= 
				\Landau{T^{\alpha}}
			\text{.}
			\label{CorrResult}
		\end{equation}
		Note that the correlation integral is $\Landau{T^{\alpha}} = \landau{T^{2\alpha}}$ and of smaller order than the denominator.
		This part does therefore not contribute in the limit $T\rightarrow\infty$.
		Hence, the overall result is given by the CTRW value:
		\begin{equation}
				J_{\text{sfBm}}
			= 
				J_{\text{CTRW}}
			= 
				2 \frac{ \Gamma^2\of{1+\alpha} }{ \Gma{1+2\alpha} } - 1  	
			\label{JResult}
		\end{equation}
		for all $\alpha \in \br{0,1}$ and $H \in \br{0,1}$.
		This is the main result of our theoretical discussion.
		The corrections to this limiting value for finite observation time $T$  are of order $\Landau{T^\alpha}$.

		As pointed out above, if $H$ is put to unity, $K\of{t_m,\tau_d}$ simplifies to $4 \xpc{ \Delta U^{2}_{t_m} \Delta U^{2}_{t_m+\tau_d} }$, and $J$ is equal to $10 \Gamma^2\of{1+\alpha}/\Gma{1+2\alpha} - 1$.
		In the framework of our approximations, this transition is discontinuous and contributes to extremely slow convergence of $J$ for $H$ close to unity.

		\begin{figure}
			\includegraphics[width=0.49\textwidth]{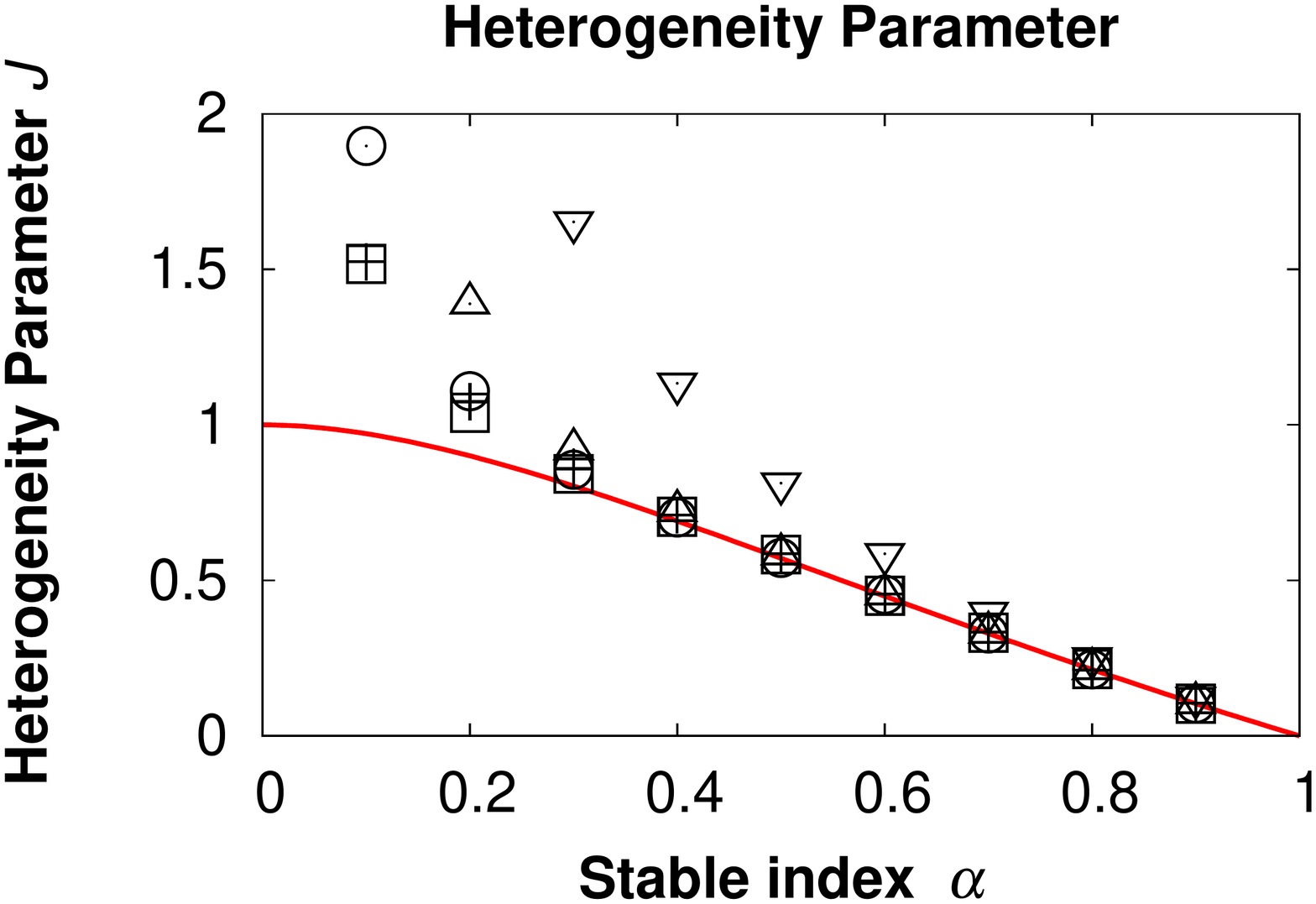}
			\caption{
				Heterogeneity Parameter.
				The plot shows the prediction of eq.\eqref{JResult} (solid line) in comparison with simulations.
				Each symbol is an average of 15 simulations with 4096 trajectories, $T = 10^6$ and $\tau = 2^{-17}T$.
				Hurst indices are $H=0.1$ ($+$), $H=0.3$ ($\square$), $H=0.5$ ($\circ$), $H=0.7$ ($\triangle$) and $H=0.9$ ($\triangledown$).
				The agreement is good for large $\alpha$ and gets worse for higher $H$ and small $\alpha$.
				\label{JPlot}
			}
		\end{figure}

		FIG.~\ref{JPlot} shows the theoretical prediction of eq.\eqref{JResult} compared with simulation data.
		Let us give a short presentation of our simulation algorithm.
		$U\of{t}$ is given by a renewal process, that means that $U\of{0} = 0$ and $U$ is increased by unity after some time $\delta t$, drawn from an one-sided $\alpha$-stable distribution, see \cite[p. 108]{Klafter2011}.
		The fBm process $\tilde{X}\of{u}$ is approximated by a Mandelbrot-Weierstrass process:
		\begin{equation}
				\tilde{X}\of{u} 
			= 
				\Sum{n=-n_{min}}{n_{max}} 
				\gamma^{-n H} \brr{
					\cos\of{\phi_n}
					-
					\cos\of{2 \pi \gamma^n u / u_{max} + \phi_n}
				}
			\text{.}
			\label{FBMSim}
		\end{equation}
		For further details see \cite{Saxton2001,Berry1980}.		
		The $\phi_n$ are random phases uniformly sampled on $[0,2\pi)$ and determine the trajectory of $\tilde{X}$.
		$\gamma$ has to be an irrational number greater than one, here chosen to be $\sqrt{\pi}$.
		$u_{max}$ was chosen to be $10 T/\tau$, ensuring that $U\of{t} < u_{max}$ during the simulation time.
		The series was evaluated from $n_{min} = -10$ to $n_{max} = 50$.

		The agreement of the data with theory is very good for moderate values of $\alpha$, and for $H$ not too close to unity, the point of transition.
		Simulations get much more difficult in this parameter regime.
		This is because simple fractional Brownian motion with Hurst index close to unity is a highly persistent process.
		Even for $\alpha = 1$, when $J$ converges to zero, the convergence was shown to be very slow \cite{Deng2009} (the decay is of order $T^{4H-4}$).

		For CTRW ($H = 1/2$), the slow convergence for small $\alpha$ was pointed out in \cite{He2008}.
		In the cases of small $\alpha$ and large $H$ both effects seem to be superimposed, leading to extremely slow convergence and to serious deviations from the prediction within the time domain accessible for simulations.
		
\begin{figure}
			\includegraphics[width=0.49\textwidth]{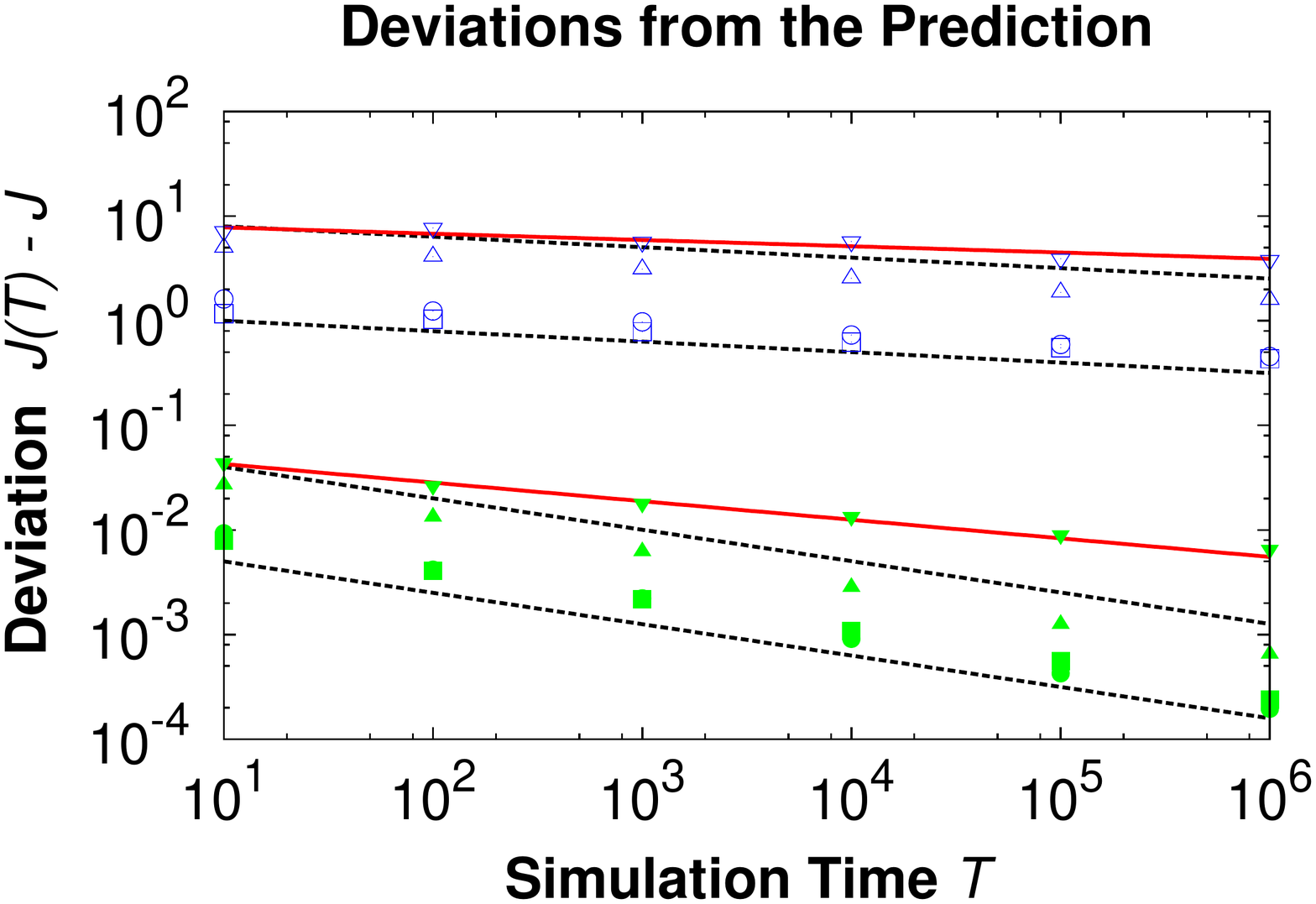}
			\caption{
				Decay of the of the deviation of $J\of{T}$ from the theoretical prediction, eq.\eqref{JResult}.
				The deviation decays as a power law $T^{-\alpha}$ except for $H=0.9$, where the decay is even slower.
				Hurst index is $0.1$ ($\square$), $0.3$ ($\odot$), $0.7$ ($\triangle$) and $0.9$ ($\triangledown$), stability parameter $\alpha$ is  $0.1$ (empty symbols) and $0.3$ (full symbols).
				The symbols for $\alpha=0.3$ are shifted for lucidity.
				For comparison dotted lines with slopes $-0.1$ (top) and $-0.3$ (bottom) are given.
				Solid lines are fits for $H=0.9$ and have the slopes $-0.18\pm0.01$ (bottom line, $\alpha = 0.3$) and $-0.06\pm0.01$ (top line, $\alpha = 0.1$).
				Each symbol is an average of $25$ simulations with $10^3$ trajectories.
				\label{JScaling}
			}
		\end{figure}

		However, for the cases, when the results in FIG.~\ref{JPlot} show large deviation from the prediction, a test of convergence can be made.
		In FIG.~\ref{JScaling} we plotted simulated values of $J\of{T} - J_{\text{sfBm}}$ against the observation time $T$.
		The data suggests, that the deviations from the limit value decay as $T^{-\alpha}$, as we estimated.
		The only exception is $H=0.9$, where the process is in the vicinity of the crossover.
		For this value of $H$, the observed pattern of decay is slower.
		For $\alpha = 0.1$ the decay is $T^{-0.06}$, and for $\alpha = 0.3$ it is $T^{-0.18}$, as obtained by fitting.
		Nevertheless, since the deviation decays, we conclude that eq.\eqref{JResult} is valid even in this regime.

	\textbf{Discussion:}
		In systems showing weak ergodicity breaking time averaged quantities remain random variables.
		This behavior can be characterized and quantified by a non-vanishing heterogeneity (``ergodicity breaking'') parameter.
		In the present paper we considered subordinated fractional Brownian motion (sfBm) being a composition of a renewal process and a fractional Brownian motion; the heterogeneity parameter is the normalized variance of the time averaged mean squared displacement.
		We show that the heterogeneity parameter of sfBm with different values of the Hurst exponent $H$ of the parent process is the same as for CTRW (corresponding to $H=1/2$) with the same value of $\alpha$.
		This stresses the fact that the heterogeneity parameter is a property determined by the (non-stationary) subordination properties of the process.
		In the language of \cite{Thiel2013}, it is a property of ``energetic disorder'' rather than of ``structural disorder''.
		In this sense weak ergodicity breaking is a phenomenon pertinent to far-from-equilibrium situations.

	\acknowledgements{The authors acknowledge financial support by DFG within IRTG 1740 research and training group project.

	\appendix
	\section{Properties of $U$ and estimation of the integrals}
		\subsection{Expectations and correlations of the underlying renewal process}
			Our analysis of the renewal process starts with the PDF of the inter-renewal times $\psi\of{t}$ distributed according to an totally skewed $\alpha$-stable law.
			The Laplace transform of $\psi\of{t}$ reads $\psi\of{s} = \int_{0}^{\infty}\d t e^{-st} \psi\of{t} = e^{-s^{\alpha}}$.
			The characteristic time scale gives the same constant factor to all results and cancels out in the end.
			It is set to $\cos^{1/\alpha}\of{\pi \alpha /2}$, see \cite[Prop. 1.2.12]{Samorodnitsky1994}.
			$\psi\of{t}$ is a function which vanishes at $t=0$, reaches a maximum at some $\tilde{t} > 0$ and then monotonically decays as a power law $t^{-1-\alpha}$ in the asymptotic limit $t\rightarrow\infty$.
			Another important function whose behavior will be needed is rate of renewal events $k\of{t}$ which is given in Laplace domain as $k\of{s} = 1/(1-\psi\of{s})$.
			With above given PDF, the rate of renewals decays as $k\of{t} \propto t^{\alpha-1}$ for large times.
			This form however shows a divergence for $t \rightarrow 0^+$, which prevents continuing this form to the small time domain.
			At small times, the behavior of the rate is as follows: By expanding the geometric series we obtain $k\of{s} \approx 1 + \psi\of{s}+ \hdots$ and therefore $k\of{t} \approx \delta\of{t} + \psi\of{t}$.
			Thus $k\of{t}$ for $t>0$ first follows the $\alpha$-stable law (which is a growing function for $t < \tilde{t}$) and then decays as a power law.
			Since $k\of{t}$ is a continuous function for $t>0$, it has to possess a maximum $k^*$ which is assumed at $t^*$.
			For the following estimates it is sufficient to put $k\of{t}$ equal to $k^*$ for $0<t<t^*$.

			The forward waiting time $\psi\of{\tau;t_1}$ is the probability to observe a renewal at $t_1+\tau$ but not in the interval $\br{t_1,t_1+\tau}$.
			If $\tau \leftrightarrow s$ and $t_1 \leftrightarrow s_1$ are the corresponding Laplace variables, then 
			\begin{equation*}
					\psi\of{s;s_1} 
				= 
					k\of{s_1} \frac{\psi\of{s} - \psi\of{s_1}}{s_1-s}
			\end{equation*}
			(see \cite[chapters 3 and 4]{Klafter2011} for details).
			The moments of $U$ have an exact representation in Laplace domain
			\begin{equation}
					\xpc{ U^{\gamma}\of{s} } 
				= 
					\frac{1-\psi\of{s}}{s} \Sum{m=0}{\infty} m^{\gamma} \psi^{m}\of{s} 
				,
				\label{eqUMomStartLaplace}
			\end{equation}
			for $\gamma > 0$.
			The factor $(1-\psi)/s \mapsto \chi\of{\tau}$ is the Laplace representation of the probability to have no renewals up to time $\tau$ and will be needed in what follows.
			The product $P_m\of{s} = \psi^m\of{s}\chi\of{s}$ is the Laplace transform of the probability to have exactly $m$ renewals in the interval $\br{0,\tau}$.
			The Laplace expression \eqref{eqUMomStartLaplace} can be inverted asymptotically, by approximating the sum with an integral and using a formula for the inverse moments of L\'evy-distributions, \cite[eq.(25.5)]{Sato1999}: 
			\begin{equation}
					\xpc{U^{\gamma}\of{t} } 
				= 
					\frac{ \Gma{1+\gamma} }{ \Gma{1+\alpha\gamma} } 
					t^{\alpha\gamma}  
				\text{.}
				\label{UMomStart}
			\end{equation}
			Following \cite{Klafter2011} the expected $\gamma$-th power of the number of renewals in any interval can be given in Laplace domain
			\begin{equation*}
				\xpc{ \Delta U^{\gamma}\of{t_1,t_1+\tau} } \mapsto \psi\of{s;s_1} \xpc{ U^{\gamma}\of{s} }
				\text{.}
			\end{equation*}
			By computing the convolution, we arrive at a hypergeometric function, whose asymptotic expansion for large $t_1/\tau$ reads
			\begin{equation}
					\xpc{ \Delta U^{\gamma}\of{t_1,t_1+\tau} } 
				= 
					\frac{ \Gma{1+\gamma} }{\Gma{\alpha} \Gma{2+\alpha \bar{\gamma}} } 
					\br{\frac{t_1}{\tau}}^{\alpha-1} \tau^{\alpha\gamma}
					\text{,}
				\label{UMom}
			\end{equation}
			with $\bar{\gamma} = \gamma-1$.
			This result is the same as eq.(15) of ref. \cite{Meroz2010}.

			In order to calculate correlations between powers of renewal numbers in different intervals, we need to consider $\bar{P}_m\of{\tau;t_1}$, the probability to have $m$ renewals in the interval $\br{0,t_1}$ and the next renewal at $t_1+\tau$.
			The rationale behind this is the following: The process $U$, being a renewal process, rejuvenates only at renewal epochs.
			In the above calculations we averaged over $\tau$, the time of next renewal (this reflects in the factor $\chi\of{s}$ in eq. \eqref{eqUMomStartLaplace}).
			We need the probability to have $m$ renewals without loosing information on the next rejuvenation.
			This can be done with $\bar{P}_m$.
			Let $\psi_m\of{\tau}$ be the inverse Laplace transform of $\psi^m\of{s}$, then 
			\begin{equation*}
					\bar{P}_m\of{\tau;t_1} 
				= 
					\sInt{0}{t_1}{t} \psi_m\of{t} \psi\of{t_1+\tau-t}
			\end{equation*}
			which transforms to 
			\begin{equation*}
					\bar{P}_m\of{s;s_1} 
				= 
					\psi^m\of{s_1} \frac{\psi\of{s_1} - \psi\of{s}}{s - s_1}  
				\text{.}
			\end{equation*}
			It is easy to confirm that $\sInt{0}{\infty}{t} \bar{P}_m\of{t;t_1} = P_m\of{t_1}$.
			By comparison of the Laplace expressions for $P_m$ and $\bar{P}_m$ and noting that $\bar{P}_m\of{\tau;0} = 0$, we obtain 
			\begin{equation}
					\bar{P}_m\of{\tau;t_1} 
				= 
					\frac{\d }{\d t_1} \sInt{0}{t_1}{t} P_m\of{t_1-t} \psi\of{\tau;t} 
				.
			\end{equation}

			Now we can calculate the correlation between the numbers of renewals in different intervals:
			\begin{align*}
				&
					\xpc{ 
						\Delta U^{\gamma}\of{t_1,t_1+t_2} 
						\Delta U^{\gamma}\of{t_1+t_2+t_3,t_1+t_2+t_3+t_4} 
					}																	 												\\
				= &
					\Sum{m=0}{\infty} m^{\gamma} \Sum{j=0}{\infty} j^{\gamma}
					\psi\of{t_2;t_1} \underset{t_2}{\ast} \bar{P}_m\of{t_3;t_2} 
					\underset{t_3}{\ast} \psi\of{t_4;t_3}
					\underset{t_4}{\ast} P_j\of{t_4}																					\\
				= &
					\Sum{m=0}{\infty} m^{\gamma} \psi\of{t_2;t_1} 
					\underset{t_2}{\ast} \bar{P}_m\of{t_3;t_2} 
					\underset{t_3}{\ast} 
					\xpc{ \Delta U^{\gamma}\of{t_3;t_3+t_4} }																			\\
				= &
					\frac{\d}{\d t_2}
					\Sum{m=0}{\infty} m^{\gamma} \psi\of{t_2;t_1} 
					\underset{t_2}{\ast} P_m\of{t_2} 
					\underset{t_2}{\ast} \psi\of{t_3;t_2}
					\underset{t_3}{\ast}
					\xpc{ \Delta U^{\gamma}\of{t_3;t_3+t_4} }																			\\
				= &
					\frac{\d}{\d t_2}
					\xpc{ \Delta U^{\gamma} \of{t_1,t_1+t_2} }
					\underset{t_2}{\ast} \psi\of{t_3;t_2}
					\underset{t_3}{\ast} \xpc{ \Delta U^{\gamma}\of{t_3,t_3+t_4} }														
				.
			\end{align*}
			The definition eq.\eqref{eqUMomStartLaplace} was used repeatedly here.
			The asterisk denotes a convolution integral over the corresponding variable.
			This double convolution can be carried out and we obtain
			\begin{align}
				\nonumber  &
					\xpc{ 
						\Delta U^{\gamma}\of{t_1,t_1+t_2} 
						\Delta U^{\gamma}\of{t_1+t_2+t_3,t_1+t_2+t_3+t_4} 
					}																	 												\\
				= &
					\br{ \frac{ \Gma{1+\gamma} }{\Gma{\alpha} \Gma{2+\alpha \bar{\gamma}} } }^2
					\br{\frac{t_1}{t_2}}^{\alpha-1} t_2^{\alpha\gamma}			
					\br{\frac{t_3}{t_4}}^{\alpha-1} t_4^{\alpha\gamma}			
					\text{.}
				\label{UCorr}
			\end{align}
			Again, lower orders of $t_1 / t_2$ and $t_3 / t_4$ have been neglected.
			Eq.\eqref{UMom} and eq.\eqref{UCorr} are the main results of this appendix and will be used in the following to calculate $J$ and estimate the corrections thereto.

		\subsection{Estimates of the integrals}
			For estimating the integrals in eq.\eqref{eqNasty}, we will exploit certain properties of $U$ like the decay of the renewal rate at longer times or the counting monotonicity.
			With counting monotonicity we mean the fact, that $\Delta U\of{t,t+\tau}$ is a non-decaying function of $\tau$ in each realization of $U$ and by monotony, $\xpc{ \Delta U^{\gamma}\of{t,t+\tau_1} } \le \xpc{ \Delta U^{\gamma} \of{t,t+\tau_2} }$, provided $\tau_1 \le \tau_2$.
			This holds exactly, since the inequality holds for every realization of $U$ and therefore on average.
			Furthermore we will make use of the decaying renewal rate $k\of{t}$.
			Since $k\of{t}$ is monotonically decreasing for $t>t^*$, the interval $\br{t_1,t_1+\tau}$ will on average contain more renewal events than the interval $\br{t_2,t_2+\tau}$, if $t^* \le t_1 < t_2$.
			Hence, $\xpc{ \Delta U^{\gamma}\of{t_1,t_1+\tau} } \ge \xpc{ \Delta U^{\gamma}\of{t_2,t_2+\tau} }$.
			However, one can show that the contribution of the small time domains to all integrals evaluated only leads to lower order corrections, and neglect the restriction $t > t^*$.

			To summarize this short digression: We allow ourselves to enlarge the intervals or to shift the intervals to the left in order to obtain upper bounds for the corresponding expectations.
		
			Let us first consider the integral:
			\begin{equation*}
				\sInt{0}{T-\tau}{\tau_d} \sInt{0}{T-\tau-\tau_d}{t_m} \xpc{ \Delta U_{t_m}^{2H} \Delta U_{t_m+\tau_d}^{2H} }
			\end{equation*}
			The correlator is easy to compute if the intervals do not overlap.
			Let us therefore split up the integral in $\tau_d$  at some arbitrary time $\tau < t_a \ll T-\tau$ (this is possible if $T$ is sufficiently large).
			$\int_0^{T-\tau}\d \tau_d \hdots = \int_0^{t_a}\d \tau_d \hdots + \int_{t_a}^{T-\tau}\d \tau_d \hdots$.
			Now we estimate the lower part of the integral
			\begin{align*}
				&
					\sInt{0}{t_a}{\tau_d} \sInt{0}{T-\tau-\tau_d}{t_m} \xpc{ \Delta U_{t_m}^{2H} \Delta U_{t_m+\tau_d}^{2H} }			\\
				\le &
					\sInt{0}{t_a}{\tau_d} \sInt{0}{T-\tau}{t_m} \xpc{ \Delta U_{t_m}^{2H} \Delta U_{t_m+\tau_d}^{2H} }					\\
				\le &
					\sInt{0}{t_a}{\tau_d} \sInt{0}{T-\tau}{t_m} \xpc{ \Delta U_{t_m}^{2H} \Delta U_{t_m}^{2H} }							\\
				\sim &
					t_a \tau^{1 - \alpha + 4\alpha H} \Int{0}{T-\tau}{t_m} t_m^{\alpha-1}												\\
				\sim & 
					\tau^{1 - \alpha + 4\alpha H} \br{T-\tau}^{\alpha} t_a 	
				= 
					\Landau{T^{\alpha}}
				.
			\end{align*}
			Between the second and third line, we used the fact, that the rate of renewals of $U$ is decreasing over time.
			After that we used eq.\eqref{UMom} to replace the expectation.

			In the upper part of the integral, the intervals do not overlap and we are allowed to use eq.\eqref{UCorr} to calculate the expectation.
			The integration over $t_m$ is easily carried out and we have:
			\begin{align*}
				&
					\sInt{t_a}{T-\tau}{\tau_d} \sInt{0}{T-\tau-\tau_d}{t_m} \xpc{ \Delta U_{t_m}^{2H} \Delta U_{t_m+\tau_d}^{2H} }		\\
				= &
					\frac{ \Gamma^2\of{1+2H} \tau^{2 + 2\alpha \bar{H}} }{ \Gma{\alpha} \Gma{1+\alpha} \Gamma^2\of{2+\alpha \bar{H}} }
					\Int{t_a}{T-\tau}{\tau_d} \br{T-\tau-\tau_d}^{\alpha} \br{\tau_d - \tau}^{\alpha-1} 								\\
				= &
					\frac{ 
						\Gamma^2\of{1+2H} \tau^{2 + 2\alpha \bar{H}} \br{T-\tau}^{2\alpha} 
					}{ 
						\Gma{\alpha} \Gma{1+\alpha} \Gamma^2\of{2+\alpha \bar{H}} 
					}
					\Int{0}{1-\frac{t_a}{T-\tau}}{x} x^{\alpha} \br{\tfrac{T-2\tau}{T-\tau} - x}^{\alpha-1} 					
					\text{.}
			\end{align*}
			In the last line we changed the variable of integration to $x = 1 - \tau_d/(T-\tau)$.
			$\bar{H}$ is an abbreviation for $2H-1$.
			The remaining integral converges to a Beta function for $T\rightarrow\infty$, so that the overall result reads:
			\begin{align}
				\nonumber &
					\sInt{0}{T-\tau}{\tau_d} \sInt{0}{T-\tau-\tau_d}{t_m} \xpc{ \Delta U_{t_m}^{2H} \Delta U_{t_m+\tau_d}^{2H} }		\\
				= &
					\frac{ 
						\Gamma^2\of{1+2H} \tau^{2 + 2\alpha \bar{H}} \br{T-\tau}^{2\alpha} 
					}{ 
						\Gma{1+2\alpha} \Gamma^2\of{2+\alpha \bar{H}} 
					}
					+ \Landau{T^{\alpha}}			
				\text{.}
			\end{align}
			This is equation \eqref{CovResult} of the main text.

			Let us turn to the correlation integral
			\begin{equation}
				\sInt{0}{T-\tau}{\tau_d} \sInt{0}{T-\tau-\tau_d}{t_m} \xpc{ K\of{t_m,\tau_d} }
				.
				\label{CorrInt}
			\end{equation}
			Again, we split up the $\tau_d$-integral at $\tau < t_a \ll T-\tau$.
			To obtain an upper bound for the lower part of the integral, we proceed in the following way:
			First, the negative contributions in the square brackets of eq.\eqref{CorrDef} are omitted, then the intervals of the remaining summands are stretched and shifted until they match each other:
			\begin{align*}
				&
					\Int{0}{t_a}{\tau_d} \Int{0}{T-\tau-\tau_d}{t_m} \xpc{ K\of{t_m,\tau_d}	}											\\
				= & 
					\Int{0}{t_a}{\tau_d} \Int{0}{T-\tau-\tau_d}{t_m} \bigg\langle \left[ \br{
						\Delta U_{t_m,t_m+\tau_d+\tau}^{2H}
						- \Delta U_{t_m,t_m+\tau_d}^{2H}
					} \right. 																											\\
				& \quad -
					\big(
						\Delta U_{t_m+\tau,t_m+\tau_d+\tau}^{2H}
						- \Abs{\Delta U_{t_m+\tau,t_m+\tau_d}}^{2H}
					\big) \big]^2 \bigg\rangle																							\\
				\le &
					\Int{0}{t_a}{\tau_d} \Int{0}{T-\tau-\tau_d}{t_m} 
					\xpc{ \brr{ \Delta U^{2H}_{t_m,t_m+\tau_d+\tau} + \Delta \Abs{U_{t_m+\tau,t_m+\tau_d}}^{2H} }^2 }					\\
				\le &
					\Int{0}{t_a}{\tau_d} \Int{0}{T-\tau-\tau_d}{t_m} 
					\xpc{ \brr{ \Delta U^{2H}_{t_m,t_m+\tau_d+\tau} + \Delta U^{2H}_{t_m,t_m+\tau_d+\tau} }^2 }							\\
				\le &
					4 \Int{0}{t_a}{\tau_d} \Int{0}{T-\tau}{t_m} 
					\xpc{ \Delta U^{4H}_{t_m,t_m+t_a+\tau} }
			\end{align*}
			What remains is an integral over a simple expectation, which can be calculated using eq.\eqref{UMom}.
			After computation, this part is shown to be of order $\Landau{T^{\alpha}}$.
			
			For the remaining part of the integral let us note, that -- assuming that $\tau_d$ is larger than $\tau$ -- $K$ is a functional of the following three random variables: $N_1$, the number of renewals in the interval $\br{t_m,t_m+\tau}$, $N_2$, the number of renewals in the interval $\br{t_m+\tau, t_m+\tau_d}$, and $N_3$, the number of renewals in the interval $\br{t_m+\tau_d, t_m+\tau_d+\tau}$, see FIG.~\ref{figIntervalSketch}.
			In terms of these three random variables $K$ reads:
			\begin{align}
					K\of{t_m,\tau_d} 
				\nonumber = &
					\left[ 
						\br{N_1+N_2+N_3}^{2H} 
						- \br{N_2+N_1}^{2H} -
					\right.																												\\
				&
					\left.
						- \br{N_2+N_3}^{2H} 
						+ N_2^{2H} 
					\right]^2
				.
				\label{CorrNDef}
			\end{align}
			To continue, we split up the $t_m$-integral in \eqref{CorrInt} at $0 < t_a' < T-\tau-\tau_d$.
			The upper part of the $t_m$-integral is the one with large values for $t_m$ as well as $\tau_d$, when process is already pretty old.
			Therefore it is safe to assume, that the smaller two of the three intervals in question only contain one or none renewal.
			It is easily verified, that $K\of{t_m,\tau_d}$ vanishes if one of $N_1$ or $N_2$ is zero.
			Hence, we put $N_1 = N_2 = 1$.
			\begin{figure}
				\includegraphics[width=0.49\textwidth]{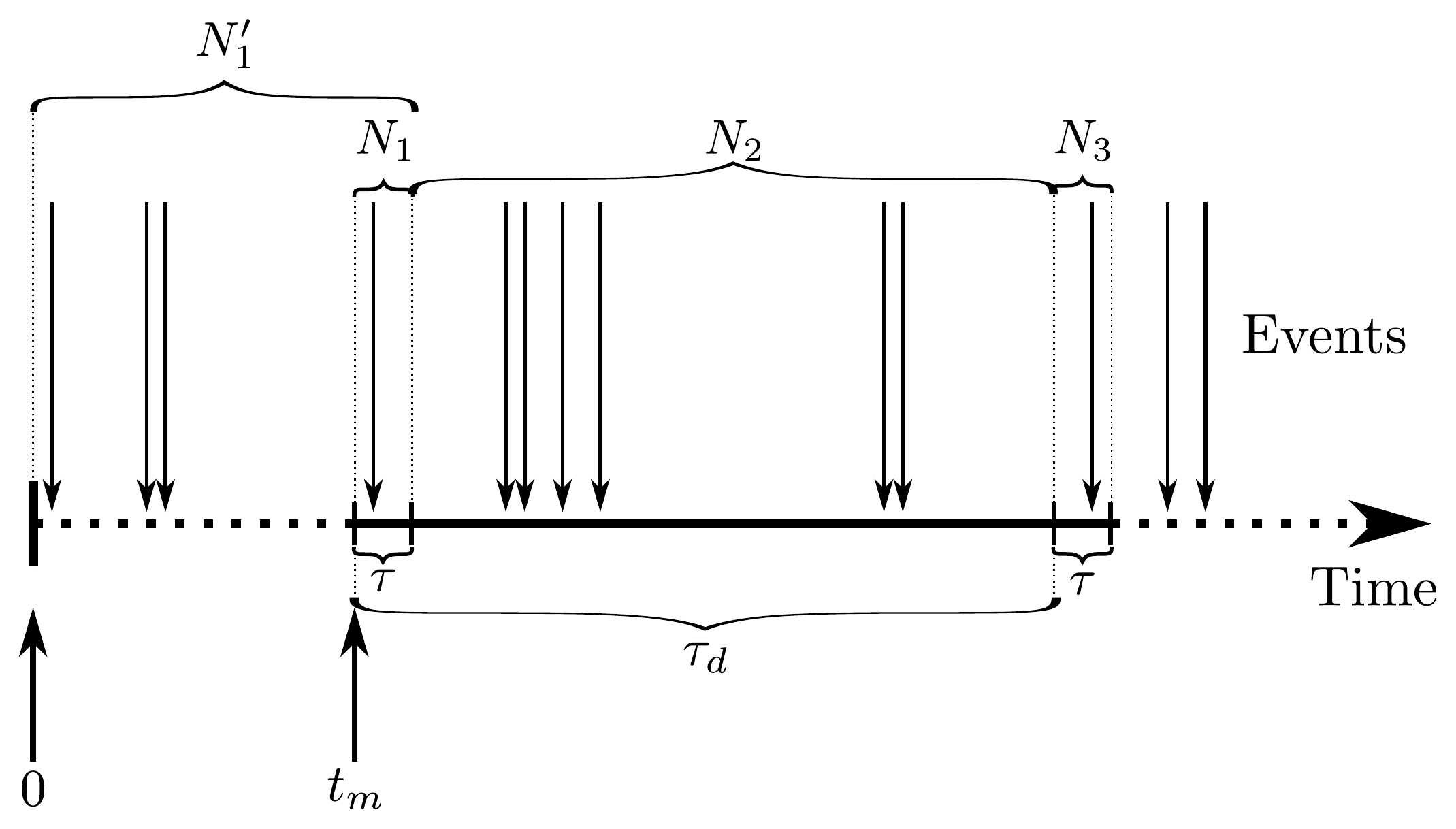}
				\caption{
					Sketch of the intervals and counting measures.
					Several renewal events (displayed as arrows directed to the bottom) occur during the evolution of $U\of{t}$.
					Below the time axis the epochs $0$ and $t_m$ are marked along with the intervals $\br{t_m,t_m+\tau}$, $\br{t_m+\tau,t_m+\tau_d}$ and $\br{t_m+\tau_d,t_m+\tau_d+\tau}$.
					The events occurring in these intervals are counted to $N_1$, $N_2$ and $N_3$.
					Since $\tau$ is very small not more than one event occurs in $\br{t_m+\tau_d,t_m+\tau_d+\tau}$.
					Events occurring after $t_m+\tau_d+\tau$ do not appear in the calculation.
					Events occurring between $0$ and $t_m+\tau$ are counted in $N_1'$.
					\label{figIntervalSketch}
				}
			\end{figure}

			We proceed by mentioning, that $\tau$ is very small w.r.t. $t_m$ or $\tau_d$, we therefore assume, that the renewal in the interval $\br{t_m,t_m+\tau}$ happens with probability $\tau k\of{t_m}$.
			A similar argument applies for the interval $\br{t_m+\tau_d,t_m+\tau_d+\tau}$.
			The probability for the event $\brrr{ N_1 = 1, N_2 = m, N_3 = 1}$ is therefore given by:
			\begin{equation*}
				\tau^2 k\of{t_m} \psi_{m+1}\of{\tau_d}
				\text{.}
			\end{equation*}
			If eq.\eqref{CorrNDef} is expanded for large $N_2$ and $k\of{t} \sim t^{\alpha-1}$ is used, the expectation reads
			\begin{equation*}
					\xpc{K\of{t_m,\tau_d}}
				\sim 
					t_m^{\alpha-1} \Sum{m=1}{\infty} m^{4H-4} \psi_{m}\of{\tau_d}
				\text{.}
			\end{equation*}
			Let us approximate this sum with an integral.
			We take $\psi\of{t}$ as an $\alpha$-stable density, then $\psi_m\of{t}$ reads $\psi_m\of{t} = m^{-1/\alpha} l_{\alpha}\of{tm^{-1/\alpha}}$.
			We use the variable transformation $x = tm^{-1/\alpha}$ and have 
			\begin{equation*}
				\Sum{m=1}{\infty} m^{4H-4} \psi_m\of{\tau_d}
				\approx 
					\alpha \tau_d^{-1 -\alpha \br{3-4H}} 
					\Int{0}{\tau_d}{x} l_{\alpha}\of{x} x^{\alpha \br{3-4H}}
			\end{equation*}
			For $H > 1/2$, the integral over $x$ converges even when the upper boundary is sent to infinity.
			We then have $\xpc{ K\of{t_m,\tau_d} } \sim t_m^{\alpha-1} \tau_d^{-1- \alpha\br{3 - 4H}}$.
			If, on the other hand, $H < 1/2$, we apply $y = x/\tau_d$ and use $l_{\alpha}\of{y\tau_d} \approx (y\tau_d)^{-1-\alpha}$.
			In this case the dependence $\xpc{ K\of{t_m,\tau_d} } \sim t_m^{\alpha-1} \tau_d^{-1-\alpha}$ is obtained.
			The average is
			\begin{equation*}
					\xpc{ K\of{t_m,\tau_d} } 
				\sim 
					t_m^{\alpha-1} \tau_d^{-1- \gamma}
				\text{,}
			\end{equation*}
			where $\gamma = \Min{\alpha}{\alpha\br{3-4H}}$.
			Integration yields:
			\begin{align*}
				&
					\sInt{t_a}{T-\tau}{\tau_d} \sInt{t_a'}{T-\tau-\tau_d}{t_m} \xpc{K\of{t_m,\tau_d}}								\\
				\sim &
					\sInt{t_a}{T-\tau}{\tau_d} \tau_d^{-1-\gamma} \br{T-\tau-\tau_d}^{\alpha}										\\
				\sim &
					T^{\alpha - \gamma} B\of{\frac{t_a}{T},1;-\gamma,1+\alpha} = \Landau{T^{\alpha}}
				,
			\end{align*}
			with the incomplete Beta integral $B\of{x,y;a,b} = \sInt{x}{y}{z} z^{a-1} (1-z)^{b-1}$.
			Note that $B\of{t_a/T,1;-\gamma,1+\alpha} = \Landau{T^{\gamma}}$.

			In the lower part of the $t_m$-integral, i.e. for $t_m < t_a'$, we can no longer assume, that the process is aged enough for $N_1=1$.
			However, since $K$ is monotonically increasing in $N_1$, we can replace it with $N_1'$, which is the number of renewals in the interval $\br{0,t_m+\tau}$.
			Since $\tau_d$ is large, we still have $N_3=1$, and $N_1' \ll N_3$.
			Substituting $N_1$ with $N_1'$ in eq.\eqref{CorrNDef} and expanding for small $N_3$ and $N_1'$ we obtain
			\begin{equation*}
					K\of{t_m,\tau_d} 
				\le 
					\br{2H}^2 \br{2H-1}^2 N_2^{4H-4} N_1'^2
			\end{equation*}
			The average over $N_1'$ and the integration over $t_m$ will result in some constant factor, since the upper bound of integration $t_a'$ is much smaller than $T-\tau$.
			Thus, we only need to care about $N_2$ and $\tau_d$ dependencies.

			Let $t'$ denote the time of last renewal before $t_m+\tau$ and $t''$ the time of first renewal after $t_m+\tau$.
			At these epochs the process is rejuvenated and the probability $I$ to have $m+1$ renewals in $\br{t_m,t_m+\tau_d}$ with the last one near $t_m+\tau_d$ is given by
			\begin{equation*}
					I 
				\approx
					\tau \sInt{t_m+\tau}{t_m+\tau_d}{t''} 
					\psi\of{t''-t'} \psi_{m}\of{t_m+\tau_d-t''} 
				,
			\end{equation*}
			conditioned on $t'$.
			Note that $\psi\of{t''-t'}$ is a monotonically decaying function of its argument, provided that $t''-t' > \tilde{t}$ and is in any case smaller than $\tilde{\psi}\of{t''-t'} = \psi^*\Theta\of{\tilde{t} - (t''-t')} + \psi\of{t''-t'}\Theta\of{t''-t'-\tilde{t}}$, where $\psi^*$ is the maximum value of $\psi\of{t}$.
			Therefore 
			\begin{align*}
					I 
				\le &
					\tau \psi^* \sInt{0}{\tilde{t}}{t''} \psi_m \of{t_m+\tau_d - t''}															\\
				&
					+ \tau \sInt{0}{t_m+\tau_d}{t''} \psi\of{t''-t'}\psi_m\of{t_m+\tau_d-t''}													\\
				\approx &
					\tau \psi_{m+1}\of{t_m+\tau_d} 
					+ \tau \tilde{t} \psi^* \psi_m\of{t_m - \tau_d - t_{mv}}
			\end{align*}
			We have enlarged the bounds of the integrals, obtaining an estimation from above for $I$.
			$t_{mv}$ is some time between zero and $\tilde{t}$ chosen according to the mean value theorem.
			Both functions appearing in this expression have the same asymptotic dependence on $\tau_d$, thus $I \sim \tau_d^{-1-\alpha}$.
			Hence, the probability for the event $\brrr{N_1'=j, N_2=m,N_3=1}$ is asymptotically proportional to $\tau_d^{-1-\alpha}$.
			Repeating the calculation from before, we have 
			\begin{equation*}
					\sInt{0}{t_a'}{t_m} \xpc{ K\of{t_m,\tau_d} } 
				\sim 
					\tau_d^{-1-\gamma}
				.
			\end{equation*}
			Since the exponent is even smaller than minus unity, the integral over the expectation gives a constant
			\begin{equation*}
					\sInt{t_a}{T-\tau}{\tau_d} \sInt{0}{t_a'}{t_m} 
					\xpc{K\of{t_m,\tau_d}}
				=
					\Landau{1}
				.
			\end{equation*}
			We therefore can summarize
			\begin{equation}
					\Int{0}{T-\tau}{\tau_d} \Int{0}{T-\tau-\tau_d}{t_m} 
					\xpc{K\of{t_m,\tau_d}} 
				=
					\Landau{T^{\alpha}}
				.
				\label{CorrEst}
			\end{equation}
			This shows eq.\eqref{CorrResult}.

	\bibliographystyle{aipnum4-1}
	\bibliography{article,book,NotRead,/home/thiel/shared/bib/self}
\end{document}